\documentstyle[11pt,twoside,pasp3D,psfig]{article}

\def\gs{\mathrel{\raise0.35ex\hbox{$\scriptstyle >$}\kern-0.6em 
\lower0.40ex\hbox{{$\scriptstyle \sim$}}}}
\def\ls{\mathrel{\raise0.35ex\hbox{$\scriptstyle <$}\kern-0.6em 
\lower0.40ex\hbox{{$\scriptstyle \sim$}}}}
\begin{document}

\title{Submm Continuum Surveys for Obscured Galaxies}
\author{Ian Smail}
\affil{Department of Physics, University of Durham, South Road, 
Durham DH1 3LE}
\author{Rob Ivison}
\affil{Department of Physics and Astronomy, University College
London, Gower Street, London WC1E 6BT}
\author{Andrew Blain}
\affil{Cavendish Laboratory, Madingley Road, Cambridge
CB3 OHE}
\author{Jean-Paul Kneib}
\affil{Observatoire Midi-Pyr\'en\'ees, CNRS-UMR5572,
14 Avenue E.\ Belin, 31400 Toulouse}
\author{Frazer Owen}
\affil{NRAO, P.O.\ Box 0, 1003 Lopezville Road, Socorro, NM 87801}

\begin{abstract}
We discuss deep surveys in the submm regime using the SCUBA bolometer
array on the JCMT.  At 850$\mu$m SCUBA has resolved the bulk of the
submm background (SMBR) detected by {\it COBE} into discrete sources
brighter than 0.5mJy. The on-going identification and characterisation
of this population at other wavelengths suggests that the bulk of the
submm sources brighter than $\sim 1$\,mJy lie at $z\gs 1$, with a
median redshift for the population of $<\! z\!  >\sim 2$--3.  The
optical/near-infrared properties of the counterparts to the submm
sources breakdown as follows: roughly two-thirds have reliable
identifications with the others being more ambiguous.  Of those with
identifications about half are optically bright ($I\ls 23$) mergers or
interactions with a high proportion showing signatures of AGN activity
and the other half are optically faint ($I\gs 25$) including both blank
fields and Extremely Red Objects (EROs).  We conclude that a population
of distant, highly obscured ultraluminous infrared galaxies (ULIRGs)
dominates the SMBR.  The wide range in the characteristics of the
optical counterparts is consistent with the dispersion in the restframe
UV properties of local ULIRGs.  We suggest that the faint submm
population comprises a class of high redshift dusty, mergers associated
with the formation of present day luminous elliptical galaxies.
\end{abstract}

%\keywords{blah}

\vspace*{-0.3cm}\section{Introduction}

The energy density in the extragalactic background radiation at
optical/UV wavelengths is roughly equal to that seen in the
far-infrared/submm (e.g.\ Bernstein et al.\ 1999; Puget et al.\ 1996;
Fixsen et al.\ 1998).  The simplest interpretation of this observation
(ignoring contributions from dust-enshrouded AGN) is that averaged over
all epochs, around half of all the star formation in the Universe has
occurred in highly obscured regions.  Clearly if we wish to obtain a
complete and unbiased view of the star formation history of the
Universe, necessary to constrain models of galaxy formation and
evolution (Baugh et al.\ 1998), we need to understand in detail what
fraction of star formation is obscured by dust and how this varies with
epoch and environment.  To achieve this we must investigate the nature
and origin of the far-infrared/submm background and that requires the
resolution of the background into discrete sources and the study of
their individual and collective properties.  

This review begins by discussing observational programmes in the submm
with the Sub-millimeter Common User Bolometer Array (SCUBA, Holland et
al.\ 1999) on the 15-m JCMT which have achieved the aim of resolving
the bulk of the extragalactic background at 850\,$\mu$m into discrete
sources.  We then discuss follow-up observations of these sources and
use these to explore the broad characteristics of the populations
contributing to the submm background.

As a benchmark for the following discussion we note that a ULIRG
similar to Arp\,220 with a far-infrared luminosity of $L_{FIR} \sim 3
\times 10^{12} L_\odot$ and a star-formation rate (SFR) of $\sim
300$\,M$_\odot$ yr$^{-1}$ would have a 850-$\mu$m flux density of $\gs
3$\,mJy out to $z\sim 10$ in a spatially flat Universe and $\gs
0.3$\,mJy for an open Universe with $q_o=0.05$ (Hughes \& Dunlop
1997).\footnote{We assume $q_o=0.5$ and $h_{\rm 100}=0.5$ unless
otherwise stated. In addition, unless identified as `observed', all
magnitudes/fluxes are corrected for lens amplification.}  In two nights
of observing in good conditions with SCUBA it is possible achieve a
3$\sigma$ flux limit of 3\,mJy across a 160$''$-diameter field, probing
a volume of 10$^6$ Mpc$^3$ out to $z\sim 10$ for dusty galaxies as
luminous as Arp\,220.

\vspace*{-0.3cm}\section{The Number Counts of the Faint Submm Population}

%
% Figure 1
%
\begin{figure}[t]
\centerline{\psfig{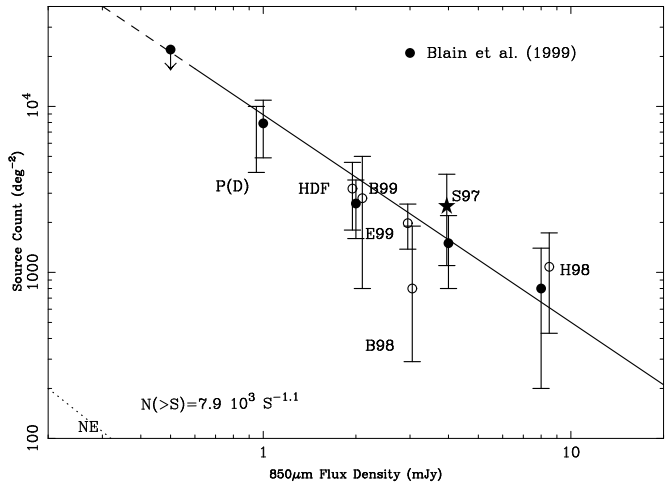}}
\caption{The cumulative 850-$\mu$m counts from published SCUBA
surveys.  The latest counts from the Blain et al.\ (1999a) analysis of
the SCUBA Cluster Lens Survey are marked by solid circles.  
Counts from Barger et al.\ (1998; B98, 1999b; B99), Eales et al.\ (1999;
E99), Holland et al.\ (1998, H98) Hughes et al.\ (1998; HDF) and Smail,
Ivison \& Blain (1997; S97) are also shown. $P(D)$ indicates the limit
from Hughes et al.'s confusion analysis in the HDF.  The solid line
shows a crude parameterisation of the counts, $N(>S)=7.9 \times 10^3
S^{-1.1}$, while the counts from a non-evolving model based on the
local {\it IRAS} 60-$\mu$m luminosity function are given by the dotted.  } 
\vspace*{-0.7cm}\end{figure}

The advent of sensitive submm imaging with SCUBA has allowed a number
of groups to undertake `blind' surveys for faint submm galaxies.
Results on the number density of sources in blank fields as a function
of limiting 850-$\mu$m flux density have been published by three
groups:  Hughes et al.\ (1998) worked with a single deep map centered
on the {\it Hubble Deep Field} (HDF); while Barger et al.\ (1998,
1999b) employed a combination of deep/narrow and wide/shallow
observations of fields in the Lockman Hole and Hawaii Survey Field
regions, finally there is a on-going survey of areas included in the
Canada-France Redshift Survey (first results given in Eales et
al.\ 1999).  The surface densities of sources measured by the different
groups are shown in Fig.~1.  Due to the modest resolution of these
maps, 15$''$ FWHM, they are confusion limited at $\sim 2$\,mJy.

Our collaboration has taken a complimentary approach to these `blank'
field surveys by using massive gravitational cluster lenses to increase
the sensitivity and resolution of SCUBA.   The first deep submm counts
were based on maps of two clusters (Smail, Ivison \& Blain 1997) and
the survey was subsequently expanded to cover seven lensing clusters at
$z=0.19$--$0.41$ (Smail et al.\ 1998; Blain et al.\ 1999a).  The
complete sample comprises a total of 17 galaxies detected at
3\,$\sigma$ significance or above, with 10 detected above 4\,$\sigma$,
from a total surveyed area of $\sim 40$\,sq.\ arcmin in the image plane
down to a $1\sigma$ flux limit of $\sim 1.5$\,mJy at 850\,$\mu$m.  Two
of these sources are identified with the central cluster galaxies in
the clusters A\,1835 and A\,2390 and as such are removed from our
analysis, although see Edge et al.\ (1999) for more discussion of these
systems.  The analysis of our catalog  makes use of well-constrained
lens models for all the clusters to accurately correct the observed
source fluxes for lens amplification (Blain et al.\ 1999a).  For the
median source amplification, $\sim 2.5\times$, our survey covers an
area of the source plane equivalent to  15 sq.\ arcmin to a $3\sigma$
flux limit of $\sim 2$\,mJy.  The lens amplification also results in a
factor of two finer beam size at this depth so that these counts have a
fainter confusion limit than the blank field observations.  At higher
amplifications, the survey covers a smaller region, but at a
correspondingly higher sensitivity (e.g.\ $\sim 1$ sq.\ arcmin at
$\sigma_{850} \sim 0.1$\,mJy) and resolution.  The uncertainties
associated with our lensing analysis are included in the final error
quoted on the derived counts (Blain et al.\ 1999a).  The total
uncertainty in the lensing correction is at most comparable to the
typical error in the absolute SCUBA calibration.  

The  850-$\mu$m counts from the analysis of Blain et al.\ (1999a) are
shown in Fig.~1 and are in agreement with the results from the other
surveys at $\geq 2$\,mJy.  However, the magnification produced by the
massive cluster lenses allows us to also constrain the source counts
down to 0.5\,mJy, four times fainter than the deepest blank-field
counts published and free from confusion noise.

The cumulative 850-$\mu$m counts down to 4\,mJy from Smail, Ivison \&
Blain (1997) accounted for roughly 30\% of the SMBR detected by {\it
COBE} (e.g.\ Puget et al.\ 1996; Fixsen et al.\ 1998).  The counts from
later surveys down to  the blank field confusion limit of JCMT at
2\,mJy account for close to 50\% of the SMBR, while the deepest counts
from the lens fields  indicate that the bulk of the SMBR is resolved by
0.5\,mJy (Blain et al.\ 1999a).  The majority of the SMBR is thus
produced by sources with 850-$\mu$m fluxes of 1--10\,mJy and as we
discuss in the next section these galaxies are likely to lie at $z\gs
1$ and hence they have intrinsic bolometric luminosities of
$10^{12}$--$10^{13} L_\odot$  and  densities of around
$10^{-5}$ Mpc$^{-3}$.    

In addition to the 850-$\mu$m maps discussed above, SCUBA also provides
simultaneous 450-$\mu$m imaging of the same fields.  However, the
combination of modest atmospheric transmission at 450\,$\mu$m in normal
conditions on Mauna Kea and the lower efficiency of the JCMT dish
surface at shorter wavelengths has restricted the results appearing in
this waveband.   Based upon a similar analysis to Blain et
al.\ (1999a), using those 450-$\mu$m maps from the lens survey where
useful sensitivity was obtained we derive an approximate cumulative
source density of 1000 deg$^{-2}$ brighter than 20\,mJy at
450\,$\mu$m.  This surface density is consistent with the number of
450-$\mu$m detections in the Eales et al.\ (1999) survey and the
reported lack of detections in the HDF by Hughes et al.\ (1998). As we
discuss in the next section the relative paucity of 450-$\mu$m sources
suggests a fairly high redshift for the bulk of the submm population,
$z\gg 1$.

Having resolved the background we can now study the nature of the
populations contributing to the SMBR and so determine at what epoch the
background was emitted.  Here again our survey has the advantage of
lens amplification, this time in the radio and optical/near-IR where the
identification and spectroscopic follow-up are undertaken.  Typically
the counterparts of our submm sources will appear $\sim 1$\,magnitude
brighter than the equivalent galaxy in a blank field.

\vspace*{-0.3cm}\section{The Redshift Distribution of the Faint Submm Population}

Photometric techniques for estimating the redshifts of candidate
counterparts to submm sources have been employed by Hughes et
al.\ (1998) and Lilly et al.\ (1999) leading to the suggestion that the
bulk of the population lay at $z=2$--4 or $z=0.1$--3 respectively.
However, these analyses are based upon the spectral energy
distributions (SED) of local optically-selected galaxies and the
diversity of the restframe UV/optical properties of local ULIRGs
(Trentham et al.\ 1999) and their differences from `normal' galaxies
suggests that such analyses are fraught with complications.  More
general constraints on the maximum possible redshift of the submm
population come from the search for the signature of Lyman-$\alpha$
absorption due to the intergalactic medium in the broad-band photometry
of candidate counterparts (Smail et al.\ 1998).  The detection of the
bulk of the proposed counterparts in our survey in $B$ or $V$ imaging
indicated that at least three-quarters were at $z\ls 5.5$ and more than
half were likely to have $z\ls 4.5$.  While admittedly weak, these
constraints are free from concerns over the  SEDs adopted for the
distant  ULIRGs and suggest that proportion of the faint submm
population at very high redshifts, $z>5$, is small (see Eales et al.\ 1999).
  
%
% Figure 2
%
\begin{figure}[t]
\centerline{\psfig{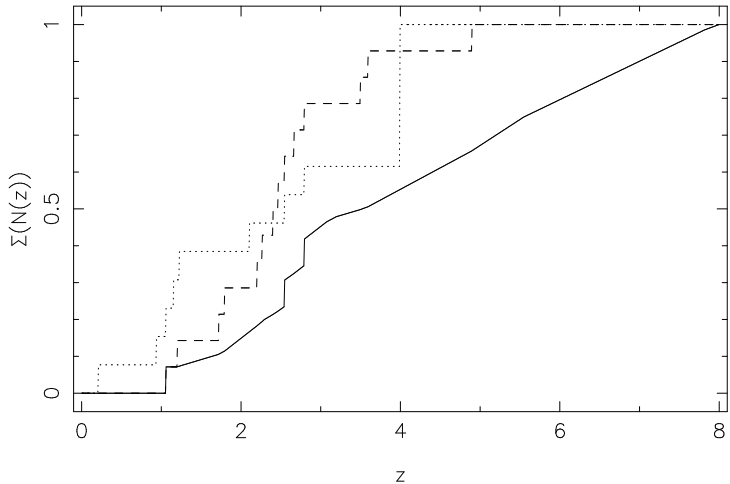}}
\caption{The cumulative redshift distribution for the SCUBA lens
survey.   We have used the spectroscopic redshifts of those galaxies
known to be reliable counterparts from Barger et al.\ (1999a) and
combined these with the probable redshift ranges of the remaining
sources derived from their observed $\alpha^{850}_{1.4}$ indices or
limits.  The solid line shows the cumulative distribution if we assume
a flat probability distribution for the sources within their allowed
$z_\alpha$ ranges from the Carilli \& Yun (1999) models and a maximum
redshift of $z=8$ for those sources where the $\alpha^{850}_{1.4}$
indices only provide a lower redshift limit.  In contrast, the dashed
line gives the conservative redshift distribution which is obtained if
{\it all} sources are assumed to lie at their lower $z_\alpha$ limit.
The dotted line is the cumulative redshift distribution for all the
counterparts from Barger et al.\ (1999a) with two of the source
identifications corrected as in Smail et al.\ (1999a) and the
blank-field/ERO candidates placed at $z=4$.} 
\vspace*{-0.5cm}
\end{figure}

First results from spectroscopic surveys of the  submm population are
beginning to appear.  In particular a Keck\,II spectroscopic survey of
possible counterparts to the submm sources in our survey has recently
been published (Barger et al.\ 1999a).  Identifications were attempted
for all the galaxies bright enough for reliable spectroscopy within the
SCUBA error-boxes.  This resulted in spectroscopic redshifts or limits
for 24 possible counterparts to 14 SCUBA sources.  The median $I$-band
magnitude of the counterparts is $I=22.4$, the equivalent depth for
identifying candidates in a blank field submm survey would be closer to
$I\sim 23$, stretching the capabilities of even the largest telescopes.  

In a number of cases the spectral properties of the candidate
counterparts suggested that they were likely to be the source of the
submm emission (Barger et al.\ 1999a), in others the identification of
a radio counterpart (e.g.\ Ivison et al.\ 1999), or the unusual
optical-NIR colors of a candidate (see \S4) or morphologies add support
to the identification of the submm emission as arising from a
particular galaxy.  For two sources we have been able to confirm the
proposed galaxy as the submm source through the detection of redshifted
CO emission in the millimeter at the redshift of the optical
counterpart (Frayer et al.\ 1998; 1999).  The reliable spectroscopic
identifications include a $z=2.8$ dusty type-2 AGN/starburst (Ivison et
al.\ 1998); a $z=2.6$ starburst (Barger et al.\ 1999a; Ivison et
al.\ 1999); a $z=3.2$ type-1 AGN (Ivison et al.\ 1999), the first
example of the sub-mJy submm population identified; and the lowest
redshift confirmed source, a $z=1.06$ ring galaxy (Soucail et
al.\ 1999).  The spectroscopic observations are thus consistent with
the photometrically-derived redshift limits for the bulk of the
population, and also provide information about the dominant emission
processes in individual galaxies. In particular, the spectra give a
useful indication of the relative fractions of AGN and starbursts in
the submm population (Ivison et al.\ 1999).  

But for over half the submm sources the results are more ambiguous
with none of the galaxies with spectroscopic identifications within the
submm error-box showing unusual spectral features, colors or
morphologies. This leaves open the possibilities that either the submm
emitting region is so highly obscured that it is invisible in the
restframe optical/UV, a not unreasonable suggestion, or that the submm
source has a fainter optical counterpart and remains unidentified in
the spectroscopic survey.  To distinguish between these alternatives
and attempt to determine the redshift distribution of a {\it
representative} sample of the faint submm population we have to resort
to other spectroscopic indicators, which while admittedly cruder have
the advantage of not relying on the identification of an optical
counterpart for the submm source.  In particular this is the only way
to tackle those sources with no visible counterparts in the optical
(e.g.\ Hughes et al.\ 1998; Smail et al.\ 1998) as well
as providing useful information on the small proportion of sources with
very red counterparts seen only in the near-infrared (Smail et
al.\ 1999a).  General constraints on the likely redshift distribution
of faint submm sources come from the spectral shape of dust emission in
the restframe far-infrared (Hughes et al.\ 1998) and the combination of
this with radio information (Carilli \& Yun 1999; Blain 1999).

As discussed by Hughes et al.\ (1998), the ratio of 450- and 850-$\mu$m
fluxes can be used as a crude redshift indicator.  Their analysis of
the information provided by the 450-$\mu$m non-detections of the five
850-$\mu$m sources  in the HDF suggested that the galaxies all
lay at $z>1$.  A similar constraint arises from assuming that the same
population of sources are being detected at 450 and 850\,$\mu$m (an
assumption which is supported by Eales et al.\ 1999) and determining
the flux ratio between the two wavelengths at a fixed source surface
density.    The 450-$\mu$m counts are 1000 deg$^{-2}$ at a flux limit
of 20\,mJy, the equivalent surface density is achieved at 850-$\mu$m at
$\sim 6.5$\,mJy.  Thus we obtain a typical $S_{450}/S_{850}$ ratio of
$S_{450}/S_{850}\sim 3$ suggesting that the median redshift for the
population is $<\! z\!> \sim 2$.

Another other method has been recently developed to estimate redshifts
for faint submm sources using the 850\,$\mu$m to 1.4\,GHz spectral index,
$\alpha^{850}_{1.4}$ (Carilli \& Yun 1999).  This technique relies upon
the good correlation between the strength of the far-infrared emission
(reprocessed UV/optical radiation from massive stars) and radio (synchtron
emission from electrons accelerated in the supernovae from massive stars)
in local star-forming galaxies (Condon 1992).  The decline in emission
from dust at longer wavelengths is eventually overtaken by the rising
synchtron emission to produce an upturn between the submm and radio
regimes around 3\,mm.  This feature is seen in the SEDs of both AGN and
starburst galaxies (see examples in Ivison et al.\ 1998, 1999) and as
proposed by Carilli \& Yun (1999) the spectral index in this region can
provide a crude redshift estimate.  The spectral index has the behaviour
that it is larger for higher redshift sources and Carilli \& Yun were able
to show that the redshift predictions from $\alpha^{850}_{1.4}$ based on
local templates spectra and model SEDs were in good agreement with the
observed redshifts for a small sample of distant submm sources.  The index
also has the useful property that contamination by radio emission from
an obscured AGN will tend to reduce the value of $\alpha^{850}_{1.4}$
giving a low redshift estimate.  Thus $\alpha^{850}_{1.4}$ can be used to
place robust {\it lower} limits on the redshifts of the submm population.

We have  used  deep VLA 1.4-GHz maps of the seven clusters in our
survey to measure or place limits on the radio flux from the submm
sources.  To make these limits as conservative as possible we have used
the radio flux for the {\it brightest} radio counterpart within each
submm error-box, irrespective of whether there are other candidates
which are preferred for other reasons.  This means that we obtain a
strong lower limit on $\alpha^{850}_{1.4}$, and hence on the source
redshift, independent of the exact source identification.  The radio
maps have a typical  $1\sigma$ sensitivity of $\ls 10\mu$Jy in the
source plane and we identify radio counterparts to around half of the
submm sources, with useful limits on the remainder.  Using the various
spectral models from  Carilli \& Yun (1999) we can transform these
$\alpha^{850}_{1.4}$ measurements and limits into redshift ranges,
$z_\alpha$, for each source.  

We plot in Fig.~2 the cumulative redshift distribution derived from
the  radio/submm spectral analysis of the sources in our survey
(Smail et al.\ 1999b).  We show two curves, the first makes the most
conservative assumption that each source without a reliable spectroscopic
identification in Barger et al.\ (1999a) lies at the minimum redshift
allowed by the observed spectral index or $3\sigma$ lower limit (typically
assuming an SED similar to Arp\,220).  The other curve assumes that the
sources have uniform probability of lying anywhere in the redshift range
allowed by their $\alpha^{850}_{1.4}$ index (with a maximum redshift
for the population of $z=8$).  These distributions are compared to that
obtained by Barger et al.\ (1999a) in their spectroscopic follow-up
of these sources (where the two blank field sources and the recently
identified ERO counterparts of Smail et al.\ (1999a) are all placed
at $z=4$).  The median redshift for the complete spectroscopic sample,
including both the reliable and possible counterparts, is $<\! z\! > \sim
2.5$ equivalent to the lower limit determined from our most conservative
spectral index analysis.  Distributing the sources in a more reasonable
manner within their allowed range for $z_\alpha$ leads to median redshifts
closer to $<\! z\! > \sim 3$--4.  

Our conclusions about the source redshifts based upon both the
$S_{450}/S_{850}$ ratio and $\alpha^{850}_{1.4}$ spectral index are
sensitive to the assumed dust temperature, $T_d$, in the sources.  For
both estimators reducing $T_d$ will allow lower median redshifts for
the population (see Blain 1999).  However, only if we force the entire
submm population to have a characteristic dust temperature less than 30\,K
can we start to push the median redshift much below $<\! z\! >=2$.
 
We conclude that a variety of constraints from the observed spectral
properties of the submm population, as well as detailed spectroscopic
observations of a small number of robustly identified sources (Ivison et
al.\ 1998, 1999; Barger et al.\ 1999a), suggest that the median redshift
of the submm population lies in the range $<\! z\! > \sim 2$--3, with
few if any luminous submm galaxies at $z\ls 1$.  The submm fluxes of all
the sources detected in published surveys lie in the range $S_{850} \sim
0.5$--$10$\,mJy, assuming they lie at $z\gs 1$ then their luminosities
are $\log_{10} L_{\rm FIR} \sim 12$--$13$ and so they all class as
ultraluminous infrared galaxies.

%
% Figure 3
%
\begin{figure}[t]
\centerline{\psfig{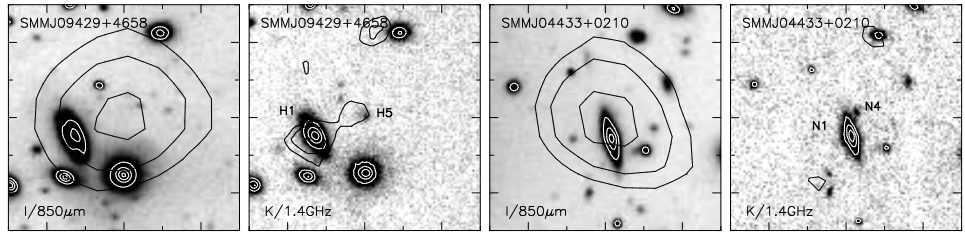}}
\caption{ERO counterparts to two submm sources in our survey (Smail et
al.\ 1999a).  The two panels on the left show a deep, 0.6$''$-resolution
Keck\,II $I$-band image and a UKIRT $K$-band image of the field of
SMM\,J09429+4658, overlayed on these are the 850-$\mu$m SCUBA map and
a deep 1.4-GHz VLA map respectively.  The two panels on the right show
the equivalent data for the field of SMM\,J04433+0210.   The faintest
sources visible in the $I$-band exposure have $I\sim 25.5$--26.0, while
the $K$-band images reach to $K\sim 20.5$.  The original candidate
counterparts for the submm sources are marked on the $K$ images, as
well as the new ERO candidates, H5 for SMM\,J09429+4658 and N4 for
SMM\,J04433+0210.  Each panel is 30$''$ square and is centred on the
nominal position of the 850-$\mu$m peak (absolute accuracy of $\ls 3''$).
The relative radio-optical astrometry is better than $0.4''$ and hence the
radio source close to the bright galaxy at the top of the SMM\,J09429+4658
frame is not coincident with it.} 
\vspace*{-0.5cm}
\end{figure}

\vspace*{-0.3cm}\section{The Nature of the Faint Submm Population}

We are still at an early stage in the study of the faint submm
population and so the following discussion will concentrate on
the results from the SCUBA lens survey for which a wide
range of follow-up has been published.  To briefly summarise the
status of this survey.  There are 15 non-cluster submm sources in
the survey,  of these around five have spectroscopic identifications
which we believe reliably identifying these galaxies as the submm
sources (two of these have been subsequently confirmed via detections
in CO).  A further three sources have either counterparts with
unambiguous radio identifications or extreme optical-near-IR colors
which allow us to identify them, and two more are in optically blank
fields.  This leaves five sources which have ambiguous identifications.   

Starting with those galaxies with spectroscopic identifications, these
all have relatively bright optical counterparts and a high proportion
show multiple components in optical/near-IR -- on separations of $\ls
2$--3$''$ (e.g.\ Ivison et al.\ 1998, 1999), at the galaxy redshifts
this scale is equivalent to $\sim 10$\,kpc.  This lends weight
to the suggestion that mergers and interactions are a crucial trigger
of activity in distant ULIRGs (Smail et al.\ 1998; Lilly et al.\ 1999)
as they are for more local examples.  The identification of submm
sources with merging systems suggests that we should properly view them as `events' rather than galaxies.

To search for any extremely red counterparts, $(I-K)>6$, which could
have been missed in the optical identifications we have used UKIRT to
obtain near-IR imaging of our fields down to $K\gs 20$.  We have so far
identified two possible ERO counterparts to submm sources in our survey
(Fig.~3, Smail et al.\ 1999a), both previously identified with bright
$z\sim 0.5$ spiral galaxies, with the bulk of the submm error-boxes
containing galaxies with optical--near-IR colors more typical of the
general field, $(I-K)\sim 2$--$4$.   Deeper $K$-band imaging of one
submm source with a reliable radio position has also provided an
identification of a $K\sim 22$ counterpart, the optical limit on this
galaxy is $I\gs 24$.  The two submm error-boxes which were `blank'
($I\gs 25$) in the optical search undertaken by Smail et al.\ (1998)
show no near-infrared candidates to $K\sim 21$.

The ERO counterparts account for 15\% of the submm population. If the
optically `blank' fields also contain EROs the proportion of submm
sources with highly reddened counterparts will rise to 30\%.  The
implied surface densities of submm-bright EROs are consistent with the
number of EROs detected in targeted SCUBA observations (e.g.\ Dey et
al.\ 1999) and suggest that around half of the ERO population are dusty, 
star forming galaxies at moderate/high redshifts (Smail et al.\ 1999a).

Although currently incomplete, we are slowly building up a picture of the
population of distant, luminous submm galaxies which dominate the SMBR
at wavelengths around 1\,mm.  The fact that the submm population detected
by SCUBA can account for all of the {\it COBE} background indicates that
a large fraction of the stars in local galaxies could be formed in these
systems.  The characteristics of the submm galaxies are similar to those
of local ULIRGs, except that they contribute a submm luminosity density
at early epochs that is at least an order of magnitude greater than the
corresponding local population.  One goal is to use this population
to trace the amount of high-redshift star-formation activity that is
obscured from view in the optical by dust, and so is missing from existing
inventories of star-formation activity at high redshift (Smail, Ivison
\& Blain 1997; Hughes et al.\ 1998; Blain et al.\ 1999b). In this way
a complete history of star formation in the Universe can be constructed.  

~From the analysis discussed in \S3 we can state that the luminous
submm population is roughly coeval with the more modestly star-forming
galaxies selected by UV/optical surveys of the distant Universe (e.g.\
Steidel et al.\ 1999).  However, the individual SCUBA galaxies have
SFRs which are typically an order of magnitude higher than those of
the optically-selected galaxies, as well as being dustier and probably
therefore more chemically enriched.  Converting the submm luminosities
into an equivalent star formation density is a very uncertain process
(e.g.\ Blain et al.\ 1999b), nevertheless, most reasonable conversions
result in a star formation density in the submm population which is a
factor of several higher than that seen in UV/optically selected samples
at high redshift (Fig.~4), even after applying similarly uncertain
corrections to the latter for the effects of dust obscuration on the
UV luminosities.   Although this comparison remains uncertain, it does
show that the submm population needs to be explained by any models that 
claim to describe the star formation history of the Universe.
  
%
% Figure 4
%
\begin{figure}[t]
\centerline{\psfig{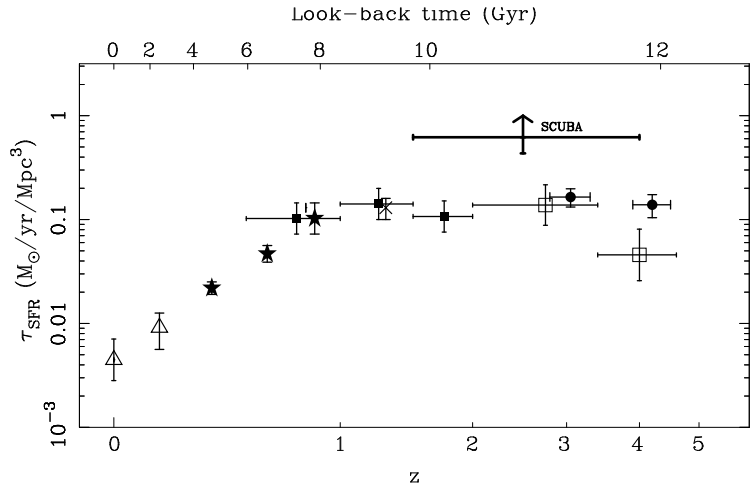}}
\caption{The estimated star formation densities at different epochs
from UV/optically-selected samples (nominally corrected for extinction)
compared to that estimated from the SCUBA population assuming the
conservative redshift distribution from Fig.~2 (Blain et al.\ 1999b).
The lower-bound on the SCUBA point shows the maximum correction for
AGN contamination in the sample as discussed in the text.  The symbols
follow those in Blain et al.\ (1999b) with the exception that the filled
circles show the latest results by Steidel et al.\ (1999) and the cross
that from Yan et al.\ (1999).}
\vspace*{-0.7cm}
\end{figure}

As with local ULIRGs, there is uncertainty over the exact contributions
from AGN and starbursts to the far-infrared luminosity density of the
SCUBA population.  This adds another possible source of contamination
to the comparison shown in Fig.~4.  The identification and removal of
obscured AGN from the submm sample  can be achieved through searches for
hard X-ray emission with {\it Chandra}.  For all but the most heavily
enshrouded systems ($\log N({\rm H}{\sc i}) \gs 24$) the hard X-ray
emission should still be detectable from the central AGN (Gunn 1999).
These searches will also provide an estimate of the total contribution
from the dust-obscured AGN to the X-ray background (Almaini et al.\ 1998;
Gunn 1999).  Calculations which use the X-ray background to constrain
the number of obscured AGN indicate that at most 20--30\% of the SCUBA
sources could harbor an obscured AGN.  The incomplete observations
available for the submm population suggest that $\gs 20$\% of the submm
population show obvious spectral signatures of an AGN. We stress, however,
that this does not mean that the AGN dominates the emission in the submm
and so the correction indicated in Fig.~4 remains uncertain.

One interesting issue which remains is what are the descendents
of  the submm population?   They have high bolometric luminosities,
large dust masses ($10^{8} M_\odot$, Ivison et al.\ 1998, 1999) and
large gas reservoirs ($10^{11} M_\odot$, Frayer et al.\ 1998, 1999).
These properties are consistent with on-going massive star formation
in these systems at rates $\gs 1000M_\odot $yr$^{-1}$, which given
the available molecular gas supply could continue for $10^8$\,yrs
and result in the formation of an entire $L^\ast$ galaxy at $z\gs
2$. The submm population also show morphological similarities to local
merging ULIRGs which are thought to evolve into elliptical galaxies.
The circumstantial evidence thus points towards the submm sources being
precursors of elliptical galaxies.  Assuming that the luminous submm phase
lasts for a few dynamical times of the remanent halo, $\gs 1$\,Gyr, the
volume density of these galaxies at high redshifts is $\sim 2$--4$\times
10^{-4}$\,Mpc$^{-3}$, high enough to allow all massive ellipticals to
be formed in this manner.  Arguments have been advanced for a number of
years for the prompt and synchronised formation of a large fraction of
the luminous elliptical galaxy populations in the richest clusters at
$z\gs 3$ (Bower, Lucey \& Ellis 1992; Ellis et al.\ 1997) due to their
extreme homogeneity both within individual clusters and between clusters.
If the SCUBA sources are identified with the earliest, obscured phases
of this activity then we would expect the submm sources to be clustered
on scales comparable to that of the putative proto-clusters at these
early epochs, $\sim 10$\,Mpc or 20 arcminutes (Governato et al.\ 1998).
Thus strong clustering of SCUBA sources is a clear prediction of the
identification of these galaxies with proto-ellipticals, searching
for such structures should be a high priority for future submm surveys.

An equally interesting issue is the relationship between the submm
population and the Lyman-break galaxies (Steidel et al.\ 1999).
These objects appear to have typically lower SFRs and less dust than the
submm galaxies, although they have a substantially higher number density.
We would therefore identify the submm galaxies with the most energetic
mergers which form massive, young ellipticals and more quiescent star
formation due to secular evolution in disk systems with the Lyman-break
objects.  Detailed observations of the submm population should thus
provide much needed observational input to models of the formation and
evolution of massive galaxies (Blain et al.\ 1999b).  In particular we
anticipate CO detections of more submm galaxies to study the kinematics
of these systems and hence determine their masses.  

\vspace*{-0.5cm}\section{The Future}

On-going and planned upgrades to SCUBA and the JCMT will improve the
sensitivity and effectiveness of this world-class facility.  Looking
slightly further ahead there is a proposal for a wide-field submm imager,
SCUBA--2, based upon new detector technology and capable of providing
statistically reliable samples of submm galaxies.  Such samples will
enable us to break new ground in crucial areas of study such as clustering
of submm sources, necessary to understand the evolutionary status of this
population relative to other classes of high redshift source.  However,
at 850\,$\mu$m the large JCMT beam will remain the main restriction to
probing deeper into the submm counts to identify the turn-over which
should occur around $\sim 0.5$--1\,mJy.  This turn-over has important
implications for models of the formation and evolution of obscured
galaxies and the approach we have taken of employing massive gravitational
lenses to increase the sensitivity and resolution of the SCUBA maps is
well suited to tackling this problem.  An equivalent length exposure to
that obtained with SCUBA on the HDF (Hughes et al.\ 1998) but instead
targetted on a well-constrained cluster lens such as Abell\,370 would
constrain the form of the 850-$\mu$m counts down to 0.3\,mJy ($3\sigma$),
probing the region where the counts should turn over if they are
to remain consistent with the {\it COBE} measurements of the SMBR.

In the long term the proposed Atacama Large Millimeter Array (ALMA) will
mark an enormous leap forward in the capabilities of ground-based submm
mapping and imaging.  The current optical/near-infrared identification
programs for the minute samples of relatively bright submm sources
provided by SCUBA are stretching the capabilities of 4- and 10-m
telescopes.  The numbers and characteristics of the sources likely to be
uncovered with ALMA will probably exceed the follow-up capabilities of
the available facilities (including {\it NGST}\,), even if we restrict
ourselves to merely near-infrared imaging and limited spectroscopy.
It may be that we have to abandon entirely short-wavelength observations
of this population and rely on what we can observe in the submm and radio.
The other option is to tailor the submm surveys for easier follow-up. In
this regard we are pursuing submm imaging of fields around bright,
$V\sim 12$, high-latitude stars (which are invisible in the submm).
These can then be used as natural guidestars for high-order adaptive
optics systems on 4- and 8-m telescopes, facilitating deep high-resolution
imaging and spectroscopy of possible counterparts.

\medskip%\acknowledgments
We are grateful to Amy Barger, Len Cowie, Katherine Gunn, Neil Trentham
for useful discussions.  IRS thanks the organisers for support to
attend the conference and acknowledges a Royal Society Fellowship.

\end{document}